\documentclass[twocolumn]{aa}
\usepackage{epsfig}

\def\amx{{$\alpha_{\mu x}$~}}

\def\ergs{{erg~cm$^{-2}$s$^{-1}$~}}

\newcommand{\lsim}{{\lower.5ex\hbox{$\; \buildrel < \over \sim \;$}}}
\newcommand{\gsim}{{\lower.5ex\hbox{$\; \buildrel > \over \sim \;$}}}

\def\ee{\end{equation}}
\def\be{\begin{equation}}

\begin{document}

\title{Swift detection of all previously undetected blazars in a micro-wave flux-limited
sample of WMAP foreground sources}
\author{P.~Giommi\inst{1}$^,$\inst{2}, M. Capalbi \inst{1}, E. Cavazzuti \inst{1}$^,$\inst{2},
S. Colafrancesco\inst{3}, A. Cucchiara\inst{4}, A. Falcone\inst{4}, J. Kennea\inst{4},R. Nesci\inst{5}, M. Perri\inst{1}, G. Tagliaferri\inst{6}, A. Tramacere \inst{5},
G. Tosti\inst{7}, A. J. Blustin\inst{8}, G. Branduardi-Raymont\inst{8}, D. N. Burrows\inst{4},  G. Chincarini \inst{6},
A. J. Dean\inst{9}, N. Gehrels\inst{10}, H. Krimm\inst{10}, F. Marshall\inst{10}, A. M. Parsons\inst{10}, B. Zhang\inst{11}
\institute{
            ASI Science Data Center, ASDC c/o ESRIN,
            via G. Galilei 00044 Frascati, Italy.
\and
            Agenzia Spaziale Italiana,
            Unit\'a Osservazione dell'Universo,
            viale Liegi, 26 00198 Roma, Italy
\and
           INAF - Osservatorio Astronomico di Roma
            via Frascati 33, I-00040 Monteporzio, Italy.
\and
           Department of Astronomy and Astrophysics,  Pennsylvania  State University, USA.
\and
            Universit\'a di Roma "La Sapienza"
            Dipartimento di Fisica
            P.le A. Moro, 2 00185, Roma, Italy
\and
            INAF - Osservatorio Astronomico di Brera
            via Bianchi 46, 23807 Merate, Italy.
\and
            Dipartimento di Fisica, Universit\'a di Perugia, Via A. Pascoli, Perugia, Italy.
\and
            UCL, Mullard Space Science Laboratory, Holmbury St. Mary, Dorking, Surrey RH5 6NT, UK.
\and
           University of Southampton, Highfield, S017 1BJ Southampton, UK.
\and
            NASA/Goddard Space Flight Center, Greenbelt, Maryland 20771, USA.
\and
           Department of Physics, University of Nevada, Las Vegas, NV 89154-4002, USA
}}
\offprints{paolo.giommi@asi.it}
\date{Received; Accepted}

\authorrunning{P. Giommi et al.}
    \titlerunning{running title}

\authorrunning{P. Giommi et al.}
    \titlerunning{Swift XRT observations of WMAP-selected blazars}

\abstract{Almost the totality of the bright foreground sources in the WMAP Cosmic
Microwave Background (CMB) maps are blazars, a class of sources that show usually also
X-ray emission. However, 23 objects in a flux-limited sample of 140 blazars of the WMAP
catalog (first year) were never reported before as X-ray sources. We present here the
results of 41 Swift observations which led to the detection of all these 23 blazars in
the 0.3-10 keV band.  We conclude that {\it all} micro-wave selected blazars are X-ray
emitters and that the distribution of the micro-wave to X-ray spectral slope (\amx) of
LBL blazars is very narrow, confirming that the X-ray flux of most blazars is a very good
estimator of their micro-wave emission. The X-ray spectral shape of all the objects that
were observed long enough to allow spectral analysis is flat and consistent with inverse
Compton emission within the commonly accepted view where the radiation from blazars is
emitted in a Sychrotron-Inverse-Compton scenario. We predict that all blazars and most
radio galaxies above the sensitivity limit of the WMAP and of the Planck CMB missions are
X-ray sources  detectable by the present generation of X-ray satellites. An hypothetical
all-sky soft X-ray survey with sensitivity of approximately $10^{-15}$ \ergs would be
crucial to locate and remove over 100,000 blazars from CMB temperature and polarization
maps and therefore accurately clean the primordial CMB signal from the largest population
of extragalactic foreground contaminants.

 }

\maketitle

\section{Introduction}

Blazars are a type of AGN whose nuclear emission is dominated by strong non-thermal
radiation  that extends across the entire electromagnetic spectrum, from radio
frequencies to the most energetic $\gamma$-rays. The typical observational properties of
blazars include irregular, rapid and often very large variability, apparent super-luminal
motion, flat radio spectrum, and large and variable polarization at radio and optical
frequencies. These unusual properties are thought to be due to electromagnetic radiation
emitted in a relativistic jet viewed closely along the line of sight
(\cite{bla78,Urry95}). There are two types of blazars: BL Lacertae objects (or BL Lacs)
that show no or very weak emission lines in their optical spectrum and Flat Spectrum
Radio Quasars (FSRQs) whose spectrum includes all the broad emission lines typically seen
in radio quiet AGN.

Traditionally, this type of sources has been found either in radio surveys or as
counterparts of X-ray sources. More recently a significant number of new blazars have
also been discovered at optical frequencies (e.g., \cite{Collinge05}).

The broad-band electromagnetic spectrum of blazars is composed of a synchrotron
low-energy component that peaks  [in a $Log(\nu f(\nu))-Log(\nu)$ representation] between
the far infrared and the X-ray band, followed by an inverse Compton component that has
its maximum in the hard X-ray band or at higher energies, depending on the location of
the synchrotron peak, and extends into the $\gamma$-ray or even the TeV energy band.
Blazars with the synchrotron peak located at infrared frequencies are usually called Low
energy peaked Blazars or LBL, while objects where the synchrotron emission reaches the
UV/X-ray band are called High energy peaked Blazars or HBL (\cite{P95}). HBLs are  almost
exclusively of the BL Lac type, whereas LBL sources are both FSRQs and BL Lac objects and
make up the large majority of the blazar population (e.g. \cite{Pad03, Gio06a}).

The micro-wave band is particularly effective for the selection of blazars, as in this
energy region the flat spectrum emission from these sources is not diluted by steep
non-nuclear components that  are often observed at cm wavelengths. However, until very
recently no surveys had sufficient  sensitivity, contrast,  and resolution to distinguish
faint foreground sources from the much stronger Cosmic Microwave Background. The
Wilkinson Microwave Anisotropy Probe (\cite{bennett03a}), which is dedicated to the
accurate measurement of the primordial fluctuations of the CMB at small angular scales,
is the first experiment to provide such a capability.  A catalog of 208 bright foreground
sources was produced by \cite{bennett03} by using the data from the first year of WMAP
data.  \cite{giocol04} have shown that the large majority of these sources ($>$85\%) are
blazars. These authors also showed that a large number of fainter, non-resolved blazars
can contaminate the primordial CMB fluctuation spectrum in a non-negligible way.

Although blazars only represent a small minority of the entire AGN population, their
strong emission at all wavelengths allow them to stand out in those energy bands where
other types of AGN are "quiet". Some of these spectral windows are presently poorly
explored but are rapidly becoming the subject of extensive studies by current and near
future satellites like WMAP, Planck (\cite{giocol04}), AGILE and GLAST  (\cite{Gio06a}).

The Swift Gamma-Ray-Burst (GRB) Explorer (\cite{Gehrels04}) is a multi-frequency, rapid
response space observatory that was launched on November 20, 2004. To fulfill its
purposes Swift carries three instruments on board: the Burst Alert Telescope (BAT,
\cite{Barthelmy05}) sensitive in the 15-150 keV band,  the X-Ray Telescope (XRT,
\cite{Burrows05}) sensitive in the 0.2-10.0 keV band, and the UV and Optical Telescope
(UVOT, \cite{Roming05}). The Swift observation program is fully dedicated to the
discovery and rapid follow up of GRBs. However, as these elusive sources explode at
random times and their frequency of occurrence is subject to large statistical
fluctuations, there are periods when Swift  is not engaged with GRB observations and can
be used for other scientific purposes. In this context we have started a program to
observe and monitor samples of blazars selected to represent different aspects of this
class of sources.

In this paper we report the observations of a sample of 23 blazars detected at micro-wave
frequencies by the WMAP satellite (\cite{bennett03a}) that had no previous X-ray
observations, and that were not detected in the Rosat all Sky Survey.  The  XRT
instrument on board Swift is particularly suitable for this project as it provides great
flexibility and sufficient sensitivity to detect and determine the spectral shape of
sources as faint as  $\approx 10^{-13}$\ergs in short exposures.

\section{Sample description}

The blazar content of the WMAP catalog of bright foreground sources (\cite{bennett03}) is
described in  \cite{giocol04} and \cite{Gio06a}. Out of the 208 sources in the catalog,
164 are previously known blazars, 13 are radio galaxies, 5 are steep spectrum radio QSOs
and 2 are starburst galaxies; 17 sources still remain unidentified. The only objects in
the WMAP catalog identified with Galactic sources are 2 planetary nebulae, whereas 5
objects have no radio counterparts at 1.4 or 5 GHz and are probably spurious. This sample
of blazars is flux limited at micro-wave frequencies and includes objects with rich
multi-frequency observational data, especially at radio, optical, and X-ray frequencies.
About 35 objects are also included in the $3^{rd}$ EGRET catalog of gamma-ray sources
(\cite{Hartman99,Gio06a}).

Within this sample, 42 objects have never been reported as X-ray sources and have no
counterpart in the ROSAT all sky X-ray survey (\cite{Vog99}). In the following, we
consider the subset of 23 X-ray undetected WMAP sources identified with blazars that have
micro-wave flux larger than 0.8 Jy at 41 GHz.
% and radio flux larger than 1.0 Jy at 5 GHz.
This sample includes less than 20\% of all the 140 identified blazars in the WMAP catalog
that satisfy the above flux limit. The list of sources is given in Tab. \ref{tab.sample}
where column 1 gives the source  name, column 2 gives the source number in the WMAP
catalog, columns 3 and 4 give the right ascension and declination for the J2000.0
equinox,  column 5 gives the redshift (when available), column 6 gives the radio flux
density  at 5GHz, column 7 the micro-wave flux density at 41 GHz from the WMAP catalog,
column 8 gives amount of Galactic N$_H$ estimated from 21 cm measurements
(\cite{Dick90}), and column 9 gives the blazar type (FSRQ or BL Lac)
\begin{table*}[h]
\caption{The sample: WMAP blazars with 41 GHz flux larger that 0.8 Jy for which no previous X-ray detection was available}
\begin{tabular}{lcllccccc}
\hline
\multicolumn{1}{c}{Blazar name} &
\multicolumn{1}{c}{WMAP } &
\multicolumn{1}{c}{R.A.} &
\multicolumn{1}{c}{Dec.} &
\multicolumn{1}{c}{Redshift} &
\multicolumn{1}{r}{Radio Flux} &
\multicolumn{1}{r}{Micro-wave} &
 \multicolumn{1}{c}{N$_H$} &
\multicolumn{1}{r}{Blazar Type} \\
 & Catalog &
\multicolumn{1}{c}{J2000.0} &
\multicolumn{1}{c}{J2000.0} &&
\multicolumn{1}{c}{5GHz, mJy} &
\multicolumn{1}{c}{Flux } &
\multicolumn{1}{c}{$x10^{20}$} &\\
&  number &&&&& 41GHz,Jy&cm$^{-2}$&\\
\multicolumn{1}{c}{(1)}&(2) &\multicolumn{1}{c}{(3)}&\multicolumn{1}{c}{(4)}&(5)&(6)&(7)&(8)&(9)\\

\hline
DW~0202+31&085&02 05 04.8&+32 12 29.1&1.466&934&1.3&6.7&FSRQ \\
PKS~0215+015& 096&  02 17 48.7 &+01 44 48.1       &1.715&   1011  &1.6& 3.3&FSRQ   \\
PKS~0511-220&127&05 13 49.1&$-$21 59 16.1&1.296&865&1.4&2.7&FSRQ\\
S5~0633+73& 087&06 39 21.9&+73 24 57.0&1.85&711&1.1&7.0&FSRQ\\
1Jy~0805$-$077  & 133&  08 08 15.4 &$-$07 51 07.9    &1.837& 1599 &1.4&9.8& FSRQ   \\
1Jy~1030+415&103&10 33 03.5&+41 16 05.1 &1.117&439&1.5&1.2&FSRQ\\
S5~1044+71&  083 &10 48 27.4&+71 43 33.9             &1.15&1900&1.1&3.5&FSRQ\\
PKS~1206$-$238 &  172& 12 09 02.3 &$-$24 06 19.0   &1.299&  1073   &1.1$^*$&7.4& BL Lac\\
1Jy~1213$-$172    &   173&12 15 46.7 &$-$17 31 45.8   &--& 1744    &1.6& 4.3&FSRQ\\
PKS~1313$-$333&  182&13 16 07.8&$-$33 38 58.9      &1.21&1093&1.7&4.8&FSRQ \\
1Jy~1406$-$076    &   203&14 08 56.4 &$-$07 52 24.9   &1.494& 1080  & 1.5 &2.8& FSRQ\\
1Jy~1424$-$418& 191 &14 27 56.2&$-$42 06 20.8        &1.522&2597&2.6&8.3&FSRQ \\
 1Jy~1548+056    &  007& 15 50 35.0 &+05 27 10.0    &1.422& 1766  &1.8& 4.3&FSRQ\\
 PKS~1725-795 &186& 17 33 40.7&$-$79 35 53.8&-& 419&1.0&7.6& FSRQ\\
S2~1848+28& 028 &18 50 27.4&+28 25 12.0               &2.56&1013&0.9&12.2&FSRQ \\
3C~395& 034&19 02 55.8&+31 59 40.9                         &0.635&1863&0.8&11.8&FSRQ\\
S4~2021+614  &  063& 20 22 06.5 &+61 36 56.8         &0.227& 2623 &1.4 &$14.0$& FSRQ\\
PKS~2209+236& 050 &22 12 05.9&+23 55 38.9          &1.125&1123&1.7&6.5&FSRQ \\
1Jy~2227$-$088& 024 &22 29 39.9&$-$08 32 53.1         &1.561&2423&2.2&4.5&FSRQ \\
1Jy~2255$-$282& 012 &22 58 05.8&$-$27 58 18.1         &0.926&2127&9.8&2.2&FSRQ \\
1Jy~2333$-$528&195&  23 36 12.0 &$-$52 36 20.8        &--& 1588 &0.8$^*$& 1.7& FSRQ \\
1Jy~2351+456& 074 &23 54 21.5&+45 53 03.1           &1.992&1127&1.7&10.5&FSRQ \\
PKS~2355$-$534 &  189& 23 57 53.2 &$-$53 11 12.8    &1.006& 1782   &1.3$^*$ & 2.0&FSRQ \\
\hline
$^*$ 33 GHz flux
\end{tabular}
\label{tab.sample}
\end{table*}

\section{Swift XRT observations}

%At the time of writing,
Swift performed a total of 41 observations for the blazars listed in Tab.
\ref{tab.sample}.

The XRT is usually operated in Auto State mode which automatically adjusts the readout
mode of the CCD detector to the source brightness, in an attempt to avoid  pile-up (see,
for details of the XRT observing modes \cite{Burrows05, Hill04}). Given the low
count-rate of our blazars most of the data were collected using the most sensitive Photon
Counting (PC) mode; Windowed Timing (WT) mode was used in some cases with very short
exposures.

We reduced the XRT data using the {\it XRTDAS} software (v1.8.0) developed at the ASI
Science Data Center (ASDC) and distributed within the HEAsoft  6.0.5 package by the NASA
High Energy Astrophysics Archive Research Center (HEASARC). We selected photons with
grades in the range 0-12 and used default screening parameters to produce level 2 cleaned
event files.

\subsection{Image analysis and X-ray fluxes}

X-ray images were accumulated from level 2 (cleaned and calibrated) event files,
accepting photons with energy between 0.3 and 10 keV. Source count-rates were estimated
using the DETECT routine within the XIMAGE.V~4.2  package. Net counts were then converted
into flux assuming a power law spectrum with photon index equal to 1.5 and setting the
amount of photoelectric absorption (N$_H$) equal to the Galactic value along the line of
sight (see Tab. \ref{tab.sample}).

Table  \ref{tab.observations} summarizes the results.  Column 1 gives the blazar name,
column 2 gives the observation date, column 3 the XRT net exposure in seconds, column 4
gives the count-rate in the 0.3-10 keV energy range, columns 5 and 6 give the observed
X-ray flux in the 0.5-2 and 2-10 keV bands respectively, and column 7 gives the
micro-wave (94~GHZ) to X-ray (1 keV) slope defined as follows:
\begin{equation}
\displaystyle {\alpha_{\mu x} = -{ Log(f_{1keV}/f_{94GHz})\over{Log(\nu_{1keV}/\nu_{94GHz})}} = -{Log(f_{1keV}/f_{94GHz})\over{6.41}} } ~,
 \label{eq.alphamux}
\end{equation}
where $f_{1keV}$ is the de-absorbed flux at 1 keV, k-corrected assuming a power law
spectral slope of 1.5, and $f_{(94GHz)}$ is the WMAP flux density at 94~GHz when
available, or an extrapolation to 94~GHz of the 41~GHz flux given in Tab.
\ref{tab.sample} assuming a spectral slope $\alpha_{\rm r} = 0.4$  ($f \propto
\nu^{-\alpha_{\rm r}}$).
All  sources were detected above the 3--sigma confidence level, with the exception of S4
2121+614, for which the X-ray fluxes is at the $\approx$ 2.1--sigma confidence level.

\begin{figure*}[ht]
 \begin{center}
 \epsfig{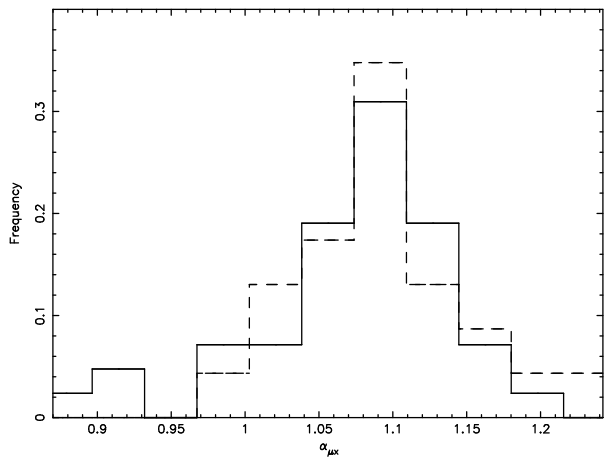}
 \end{center}
 \caption{The observed distributions of
broad-band spectral slopes between the micro-wave (94~GHZ) and X-ray (1keV) bands of the
sample of blazars detected in the 94~GHz WMAP channel and for which X-ray data were
available before our Swift observations (solid line, adapted from \cite{Gio06a}) and of
the sample presented here (dashed line).}
 \label{amx}
\end{figure*}

The X-ray fluxes of columns 5 and 6 are fairly low, but typical of LBL blazars with
similar radio fluxes. We  verified this by comparing the distribution of the $\alpha_{\mu
x}$ values listed in column 7 of Tab. \ref{tab.sample} to the distribution of the same
parameter in the sample of WMAP blazars detected both at 94~GHZ and in the X-ray band
considered by \cite{Gio05} (see  Fig.\ref{amx}). All values of  $\alpha_{\mu x}$  in Tab.
\ref{tab.sample} are within the range observed in the sample of \cite{Gio06a}, the two
distributions (solid and dashed lines in Fig. \ref{amx}) are similar, and the average
value of $1.10\pm0.2$ is consistent with the average value (1.07) in the sample of
\cite{Gio06a}.

\subsection{Spectral analysis}

Spectral deconvolution can be performed only on those objects that were observed long
enough to allow the detection of at least 130 photons in single or (when possible) merged
observations. Table 3 list the subset of the XRT observations reported in Tab.2 for which
we have found more than 130 photons in single and/or merged observations; the dates for
the considered observation are reported in column 2 of Tab.3. The photon indices for the
single and merged observations are reported in col.3 of this table and are labelled with
the suffix $(^a)$ when the spectra are obtained from merged observations.
Spectra were extracted from a circular region around the source with radius of 20 pixels,
which includes 90\% of the source photons. In order to use $\chi^2$ statistics, spectra
were rebinned to include at least 20 photons in each energy channel. We used the
XSPEC11.3 package and, given the very limited statistics, we only fitted the data to a
simple power law spectrum with low energy absorption ($N_H$) fixed to the Galactic value
in the direction of each source.  A 5\% systematic error was included to take into
account calibration uncertainties (\cite{Campana06}). The results are reported in Tab.
\ref{tab.spectral}  where column 1 gives the blazar name, column 2 gives the observation
date,  column 3 gives the best fit photon index with one sigma errors, column 4 the
reduced $\chi^2$  and column 5 the number of degrees of freedom.

Best fit spectra were de-reddened using the cross sections of \cite{Mor83} and included
in the Spectral Energy Distributions (SEDs) shown in Figs. \ref{seds_1} through
\ref{seds_3} that were built with non-simultaneous multi-frequency data taken from NED
and other catalogs available at the ASDC.

For the objects for which we collected between 50 and 130 photons in each single
observation listed in Tab.2, a (power law) spectral slope was estimated from the hardness
ratio between the 0.5-2.5 keV and the 2.5-10 keV band. In this case, no observation
merging has been adopted in the spectral analysis. Again the amount of $N_H$ was assumed
to be equal to the Galactic value. The results are reported in Tab. \ref{tab.HR} where
column 1 gives the source name, column 2 the observation date, columns 3 and 4 the
count-rates in the 0.5-2.5 and 2.5-10.0 keV bands, column 5 gives the (photon) spectral
index derived from the hardness ratio. In all cases the spectral slopes are flat,
consistent with the values obtained from the brighter sources, and indicative of inverse
Compton emission.

\begin{table*}[h]
\caption{XRT observations and results of X-ray image analysis on data taken in Photon Counting mode}
\begin{tabular}{lrcrclccc}
\hline
\multicolumn{1}{c}{Blazar name} &
\multicolumn{1}{c}{Obs. date} &
\multicolumn{3}{c}{XRT exposure} &
\multicolumn{1}{c}{XRT Count Rate} &
 \multicolumn{1}{c}{X-ray Flux} &
  \multicolumn{1}{c}{X-ray Flux} &
 \multicolumn{1}{c}{ \amx } \\
& &
\multicolumn{3}{c}{seconds}&
\multicolumn{1}{c}{0.3-10. keV} &
 \multicolumn{1}{c}{0.5-2 keV}&
 \multicolumn{1}{c}{2-10 keV} \\
& & & & &
\multicolumn{1}{c}{cts/s}&  \multicolumn{1}{c}{\ergs}& \multicolumn{1}{c}{\ergs}\\
\multicolumn{1}{c}{(1)}&\multicolumn{1}{c}{(2)}&\multicolumn{3}{c}{(3)}&\multicolumn{1}{c}{(4)}&(5)&(6)&(7)\\
\hline
DW~0202+31&24-Jun-2006& &11844& &(2.48$\pm$0.16)$\times10^{-2}$&$3.2\times 10^{-13}$&$9.3\times 10^{-13}$&1.08\\
PKS~0215+015&28-Jun-2005&&9360&&(6.0$\pm$0.30)$\times10^{-2}$&$7.8\times 10^{-13}$&$2.1\times 10^{-12}$&1.04\\
PKS~0511-220&30-Mar-2006&&2987&&(1.36$\pm$0.23)$\times10^{-2}$&$5.8\times 10^{-13}$&$1.5\times 10^{-12}$&1.05\\
                            &06-Apr-2006&&2781&&(1.56$\pm$0.26)$\times10^{-2}$&$2.2\times 10^{-13}$&$2.20\times 10^{-13}$&1.12\\
                            &10-Apr-2006&&4405&&(9.6$\pm$1.6)$\times10^{-3}$&$1.34\times 10^{-13}$&$1.34\times 10^{-13}$&1.15\\
                            &20-Apr-2006&&1029&&(1.2$\pm$0.4)$\times10^{-2}$&$1.68\times 10^{-13}$&$1.68\times 10^{-13}$&1.13\\
S5~0633+73&09-Apr-2006&&11432&&(2.07$\pm$0.15)$\times10^{-2}$&$2.69\times 10^{-13}$&$7.87\times 10^{-13}$&1.09\\
1Jy~0805$-$077 &22-May-2005&&5847&&(2.4$\pm$0.2)$\times10^{-2}$&$3.1\times 10^{-13}$&$9.6\times 10^{-13}$&1.09\\
                               &26-May-2005&&916&&(2.5$\pm$0.6)$\times10^{-2}$&$3.2\times 10^{-13}$&$1.0\times 10^{-12}$&1.09\\
                               &04-Jun-2005&&12955&&(2.6$\pm$0.2)$\times10^{-2}$&$3.3\times 10^{-13}$&$1.0\times 10^{-12}$&1.09\\
                               &11-Sep-2005&&4578&&(2.3$\pm$0.3)$\times10^{-2}$&$2.9\times 10^{-13}$&$9.2\times 10^{-13}$&1.10\\
                               &25-Sep-2005&&9016&&(2.1$\pm$0.2)$\times10^{-2}$&$2.7\times 10^{-13}$&$8.4\times 10^{-13}$&1.10\\
                               &01-Oct-2005&&3963&&(1.9$\pm$0.3)$\times10^{-2}$&$2.4\times 10^{-13}$&$7.6\times 10^{-13}$&1.11\\
1Jy~1030+415&21-Feb-2006&&3088&&(2.17$\pm$0.30)$\times10^{-2}$&$2.91\times 10^{-13}$&$6.38\times 10^{-13}$&1.10\\
                           &30-Jun-2006&&8746&&(2.14$\pm$0.18)$\times10^{-2}$&$2.87\times 10^{-13}$&$6.31\times 10^{-13}$&1.10\\
S5~1044+71&16-Apr-2005&&1770&&(4.3$\pm$0.5)$\times10^{-2}$&$5.6\times 10^{-13}$&$1.5\times 10^{-12}$&1.04\\
                        &08-Jun-2005&&732&&(3.9$\pm$0.8)$\times10^{-2}$&$5.1\times 10^{-13}$&$1.4\times 10^{-12}$&1.04\\
                        &16-Nov-2005&&2159&&(3.0$\pm$0.4)$\times10^{-2}$&$3.9\times 10^{-13}$&$1.1\times 10^{-13}$&1.06\\
PKS~1206$-$238&17-Dec-2005&&5744&& (1.7$\pm$0.2)$\times10^{-2}$&$2.2\times 10^{-13}$&$5.8\times 10^{-13}$&1.10 \\
1Jy~1213$-$172&17-Dec-2005&&10282&& (2.5$\pm$0.2)$\times10^{-2}$&$3.3\times 10^{-13}$&$9.0\times 10^{-13}$&1.10 \\
PKS~1313$-$333 &12-Jul-2005&&298&& (5.5$\pm$1.5)$\times10^{-2}$&$7.2\times 10^{-13}$&$2.0\times 10^{-12}$& 1.05\\
                                  &07-Sep-2005&&6358&& (4.9$\pm$0.3)$\times10^{-2}$&$6.4\times 10^{-13}$&$1.8\times 10^{-12}$& 1.06\\
1Jy~1406$-$076 &04-Sep-2005&&26457&&(1.9$\pm$0.1)$\times10^{-2}$&$2.5\times 10^{-13}$&$6.6\times 10^{-13}$&1.11 \\
1Jy~1424$-$418 &19-Apr-2005&&2269&&(5.6$\pm$0.5)$\times10^{-2}$&$7.2\times 10^{-13}$&$2.2\times 10^{-12}$&1.08 \\
                                &23-Apr-2005&&1549&&(3.3$\pm$0.5)$\times10^{-2}$&$4.3\times 10^{-13}$&$1.3\times 10^{-12}$&1.11 \\
1Jy~1548+056  &15-Sep-2005&&5732&&(4.1$\pm$0.3)$\times10^{-2}$&$5.3\times 10^{-13}$&$1.5\times 10^{-12}$&1.07\\
                             &30-Sep-2005&&4445&&(4.7$\pm$0.4)$\times10^{-2}$&$6.1\times 10^{-13}$&$1.7\times 10^{-12}$&1.0\\
PKS~1725-795&04-Feb-2006&&2721&&(5.0$\pm$0.5)$\times10^{-2}$&$6.5\times 10^{-13}$&$1.9\times 10^{-12}$&1.02\\
                            &20-Apr-2006&&3418&&(6.3$\pm$0.5)$\times10^{-2}$&$8.1\times 10^{-13}$&$2.4\times 10^{-12}$&1.01\\
S2~1848+28&17-May-2005&&4546&&(1.1$\pm$0.2)$\times10^{-2}$&$1.4\times 10^{-13}$&$4.6\times 10^{-13}$&1.11\\
                       &18-Dec-2005&&5837&&(1.9$\pm$0.2)$\times10^{-2}$&$2.4\times 10^{-13}$&$7.9\times 10^{-13}$&1.08\\
3C~395&02-Apr-2005&&5015&&(2.8$\pm$0.3)$\times10^{-2}$&$3.5\times 10^{-13}$&$1.2\times 10^{-12}$&1.05\\
S4~2021+614  & 06-Feb 2006 &&2324&&(2.7$\pm$1.3)$\times10^{-3}$&$3.4\times 10^{-14}$&$1.2\times 10^{-13}$&$1.25$\\
PKS~2209+236&27-Apr-2005&&1215&&(1.1$\pm$0.3)$\times10^{-2}$&$1.4\times 10^{-13}$&$4.1\times 10^{-13}$&1.16 \\
1Jy~2227$-$088&20-Apr-2005&&614&&(2.0$\pm$0.2)$\times10^{-1}$&$2.6\times 10^{-12}$ &$7.2\times 10^{-12}$&0.98\\
                               &28-Apr-2005&&9249&&(9.0$\pm$0.4)$\times10^{-2}$& $1.2\times 10^{-12}$&$3.2\times 10^{-12}$&1.03\\
1Jy~2255$-$282&23-Apr-2005&&4313&&(4.1$\pm$0.3)$\times10^{-2}$&$5.3\times 10^{-13}$&$1.4\times 10^{-12}$&1.19\\
1Jy~2333$-$528&25-Nov-2005&&8991&&(1.1$\pm$0.1)$\times10^{-2}$&$1.4\times 10^{-13}$&$3.7\times 10^{-13}$&1.11\\
1Jy~2351+456&17-May-2005&&7930&&(7.1$\pm$1.1)$\times10^{-3}$&$9.1\times 10^{-14}$&$2.9\times 10^{-13}$&1.19\\
                           &17-Aug-2005&&1933&&(1.3$\pm$0.3)$\times10^{-2}$&$1.7\times 10^{-13}$&$5.3\times 10^{-13}$&1.15\\
PKS~2355$-$534&17-Aug-2005&&4147&&(8.4$\pm$0.5)$\times10^{-2}$&$1.1\times 10^{-12}$&$2.9\times 10^{-12}$&1.00\\
\hline
\end{tabular}
\label{tab.observations}
\end{table*}

\bigskip

\begin{table*}[ht]
\caption{Results of XSPEC spectral analysis. Fits to simple power law model. }
\begin{tabular}{lcccc}
\hline
\multicolumn{1}{c}{Blazar name} &
\multicolumn{1}{c}{Observation date} &
\multicolumn{1}{c}{Photon index} &
\multicolumn{1}{c}{$\chi^2_{red}$} &
 \multicolumn{1}{r}{d.o.f.}  \\
 & & & &\\
\multicolumn{1}{c}{(1)}&\multicolumn{1}{c}{(2)}&\multicolumn{1}{c}{(3)}&\multicolumn{1}{c}{(4)}&(5)\\
\hline
 DW~0202+31    &24-Jun-2006&1.6$\pm$0.1      &1.26 &11 \\
 PKS~0215+015  &28-Jun-2005&1.53$\pm$0.07    &0.98 &20 \\
 PKS~0511$-$220&30-Mar-2006&1.8$\pm$0.2$^{a}$&0.4  &4  \\
               &06-Apr-2006&&& \\
               &10-Apr-2006&&& \\
               &20-Apr-2006&&& \\
 S5~0633+73    &09-Apr-2006&1.5$\pm$0.       &0.54 &8  \\
 1Jy~0805$-$077&22-May-2005&1.2$\pm$0.2      &1.68 &3  \\
               &04-Jun-2005&1.6$\pm$0.1      &0.46 &10 \\
               &25-Sep-2005&1.6$\pm$0.2      &1.37 &5  \\
 1Jy~1030+415  &21-Feb-2006&1.6$\pm$0.1$^{a}$&0.72&9   \\
               &30-Jun-2006&&&\\
 1Jy~1213$-$172&17-Dec-2005&1.8$\pm$0.1      &1.8 &8   \\
 PKS~1313$-$333&07-Sep-2005&1.8$\pm$0.2      &1.03&9   \\
 1Jy~1406$-$076&04-Sep-2005&1.6$\pm$0.1&0.79&17        \\
 1Jy~1548+056&15-Sep-2005&1.7$\pm$0.2       &1.93&6\\
%                           &30-Sep-2005&1.8$\pm$0.2&1.37&5\\
 PKS~1725$-$795&04-Feb-2006&1.6$\pm$0.1$^{a}$ &0.8&13 \\
               &20-Apr-2006&&&\\
 3C~395$^{*}$&02-Apr-2005&1.6$\pm$0.4&1.53&4\\
 3C~395$^{**}$&02-Apr-2005&1.7$\pm$0.2&1.47&18\\
 1Jy~2227$-$088&28-Apr-2005&1.54$\pm$0.08&1.09 &29\\
 1Jy~2255$-$282&23-Apr-2005&1.7$\pm$0.3&0.77 &4 \\
 PKS~2355$-$534&17-Aug-2005&1.5$\pm$0.2&0.53 &4 \\
\hline
\multicolumn{4}{l}{$^{a}$ Analysis performed on data merged from all observations}\\
\multicolumn{4}{l}{$^{*}$ Photon Counting data, $^{**}$ Windowed timing data (exposure of 6185 s)}
\end{tabular}
\label{tab.spectral}
\end{table*}

\begin{table*}[h]
\caption{Results of hardness-ratio spectral analysis assuming  a simple power law model. }
\begin{tabular}{lcccc}
\hline
\multicolumn{1}{c}{Blazar name} &
\multicolumn{1}{c}{Observation date} &
\multicolumn{1}{c}{Count rate}&
\multicolumn{1}{c}{Count rate} &
\multicolumn{1}{c}{Photon index} \\
& &\multicolumn{1}{c}{0.5-2.5 keV}
 &\multicolumn{1}{c}{2.5-10 keV}
 &\\
\multicolumn{1}{c}{(1)}&\multicolumn{1}{c}{(2)}&\multicolumn{1}{c}{(3)}&\multicolumn{1}{c}{(4)}&\multicolumn{1}{c}{(5)}\\
\hline
S5 1044+71&16-Apr-2005&(2.9$\pm$1.4)$\times10^{-2}$&(1.0$\pm$0.3)$\times10^{-2}$&1.6$\pm$0.4 \\
PKS 1206-238&17-Dec-2005&(1.2$\pm$0.2)$\times10^{-2}$&(3.7$\pm$0.9)$\times10^{-3}$&1.7$\pm$0.2 \\
1Jy 1424-418&19-Apr-2005&(3.7$\pm$0.4)$\times10^{-2}$&(1.9$\pm$0.3)$\times10^{-2}$&1.4$\pm$0.2 \\
                          &23-Apr-2005&(1.9$\pm$0.4)$\times10^{-2}$&(1.2$\pm$0.3)$\times10^{-2}$&1.2$\pm$0.3 \\
S2~1848+28&17-May-2005&(7.9$\pm$1.4)$\times10^{-3}$&(2.5$\pm$0.8)$\times10^{-3}$&1.8$\pm$0.3 \\
&18-Dec-2005&(1.3$\pm$0.2)$\times10^{-2}$&(5.7$\pm$1.2)$\times10^{-3}$&1.5$\pm$0.2 \\
1Jy~2333$-$528&25-Nov-2005&(7.4$\pm$1.0)$\times10^{-3}$&(2.8$\pm$0.6)$\times10^{-3}$&1.5$\pm$0.2 \\
1Jy~2351+456&17-May-2005&(4.5$\pm$0.8)$\times10^{-3}$&(2.5$\pm$0.6)$\times10^{-3}$&1.4$\pm$0.2 \\
\hline
\end{tabular}
\label{tab.HR}
\end{table*}
\begin{table*}[ht]
\caption{Results of UVOT observations. }
\begin{tabular}{lcccc}
 \hline
 \multicolumn{1}{c}{Blazar name} & \multicolumn{1}{c}{Observation date} & \multicolumn{1}{c}{Filter} & \multicolumn{1}{c}{mag} \\
 & & &\\
\multicolumn{1}{c}{(1)}&\multicolumn{1}{c}{(2)}&\multicolumn{1}{c}{(3)}&\multicolumn{1}{c}{(4)}\\
 \hline
 DW~0202+31    &24-Jun-2006 & W1  & $17.16$  \\
               &            & M2  & $17.77$  \\
               &            & W2  & $18.53$  \\
 PKS~0215+015  &28-Jun-2005 & W2  & $17.73$  \\
 PKS~0511$-$220&06-Apr-2006 & W1  & $20.22$  \\
 S5~0633+73    &09-Apr-2006 & W1  & $18.09$  \\
               &            & W2  & $19.28$  \\
 1Jy~0805$-$077&22-May-2005 & W1  & $16.42$  \\
               &            & M2  & $16.07$  \\
               &            & W2  & $18.87$  \\
 1Jy~1030+415  &21-Feb-2006 & W1  & $17.40$  \\
               &            & W2  & $17.98$  \\
 PKS~1206$-$238&17-Dec-2005 & W1  & $18.50$  \\
               &            & M2  & $19.28$  \\
               &            & W2  & $18.98$  \\
 PKS~1313$-$333&07-Sep-2005 & W1  & $17.97$ \\
               &            & M2  & $18.24$  \\
               &            & W2  & $18.43$  \\
 PKS~1725$-$795&04-Feb-2006 & W1  & $17.06$ \\
               &            & W2  & $17.44$  \\
 1Jy~2227$-$088&28-Apr-2005 & W1  & $16.94$ \\
               &            & M2  & $17.92$  \\
               &            & W2  & $17.76$  \\
 1Jy~2255$-$282&23-Apr-2005 & M2  & $16.72$ \\
               &            & W2  & $16.86$  \\
 1Jy~2351+456  &17-May-2005 & W1  & $17.67$ \\
               &            & M2  & $18.49$  \\
               &            & W2  & $18.41$  \\
 \hline
% \multicolumn{4}{l}{$^{a}$ Analysis performed on data merged from all observations}\\
%
%\multicolumn{4}{l}{$^{*}$ Photon Counting data, $^{**}$ Windowed timing data (exposure of 6185 s)}
\end{tabular}
\label{tab.uvot}
\end{table*}

\subsection{X-ray flux variability}

We searched for time variability both within single observations and in between separate
exposures whenever a source was observed more than once.\\
No significant rapid variability (i.e. within  single exposures) was detected in any
source. Some marginal evidence for flux variability within a factor of $2.2\pm 0.2$, and
$1.8\pm 0.3$ was only found between separate observations of the sources 1Jy 2227$-$088
and 1Jy 1424$-$418.

\section{Swift UVOT observations}

The Swift Ultraviolet  and Optical Telescope (UVOT \cite{Roming05}) is a 30 cm telescope
equipped with two grisms and six broadband filters.  UVOT is normally operated taking
exposure with all filters during each Swift observation, unless a very bright (mV$ <$ 12)
object is present in the field of view.

The data of all WMAP sources that were sufficiently bright to be detected by UVOT in at
least one filter were reduced with  the SWIFTTOOLS software available within the standard
HEASOFT software package.

Photometry  was performed  using a 6" aperture radius for the V, B and U filters and 12"
radius for the UVW1, UVM2 and UVW2 filters.  The raw counts  extracted in each aperture
after correction for the coincidence loss (similar to photon pile-up in X-ray CCDs), were
converted into standard magnitudes using the latest UVOT in-orbit zeropoint values (see
eg. Roming et al. 2005a  and reference therein for a full discussion about UVOT
calibration procedures and photometric accuracy in each filter).

The magnitude were de-reddened using the maps of \cite{schlegel1998} and the extinction
curve of \cite{Car89} for the optical filters (V, B and U) and those of \cite{seaton1979}
for the UV filters (UVW1, UVM2 and UVW2A). The extinction corrected magnitudes were then
converted to fluxes using the convertion factors included in the UVOT calibration data.

All useful data have been plotted as part of the SEDs shown in Figs. \ref{seds_1} through
\ref{seds_3}; the results of the UVOT observations of the brightest blazars are also
reported in Tab. \ref{tab.uvot}, where we report the Blazar name (col.1), the UVOT
observation date (col.2), the UVOT filter (col.3) and the observed magnitude (col.4).

\section{Summary and Discussion}

We have presented in this paper the results of 41 Swift observations of all the 23
micro-wave-selected blazars that were never reported as X-ray emitters in a 41 GHz
flux-limited sample of 140 blazars detected as bright WMAP foreground sources.

The main motivations that prompted this study are: {\it i}) establish the X-ray
properties of a statistically representative sample of micro-wave selected blazars that
are known to contaminate in a non-negligible way CMB fluctuation maps, {\it ii}) verify
or exclude the existence of a  population of X-ray quiet blazars, {\it iii}) compare the
X-ray properties of these sources with those of blazars selected in different energy
bands, and {\it iv}) complete the X-ray measurements of a micro-wave flux-limited sample
of blazars that is useful for multi-frequency statistical studies.

We found that all blazars were clearly detected by the Swift X-Ray Telescope even in very
short exposures. The observed 2-10 keV fluxes range between $2.2\times 10^{-13}$ and
$5.1\times 10^{-12}$ \ergs and are well within the expectations for LBL type blazars with
radio flux densities close to 1 Jy.  We therefore conclude that  there is  no evidence
for a population of X-ray quiet blazars.

\begin{figure*}[ht]
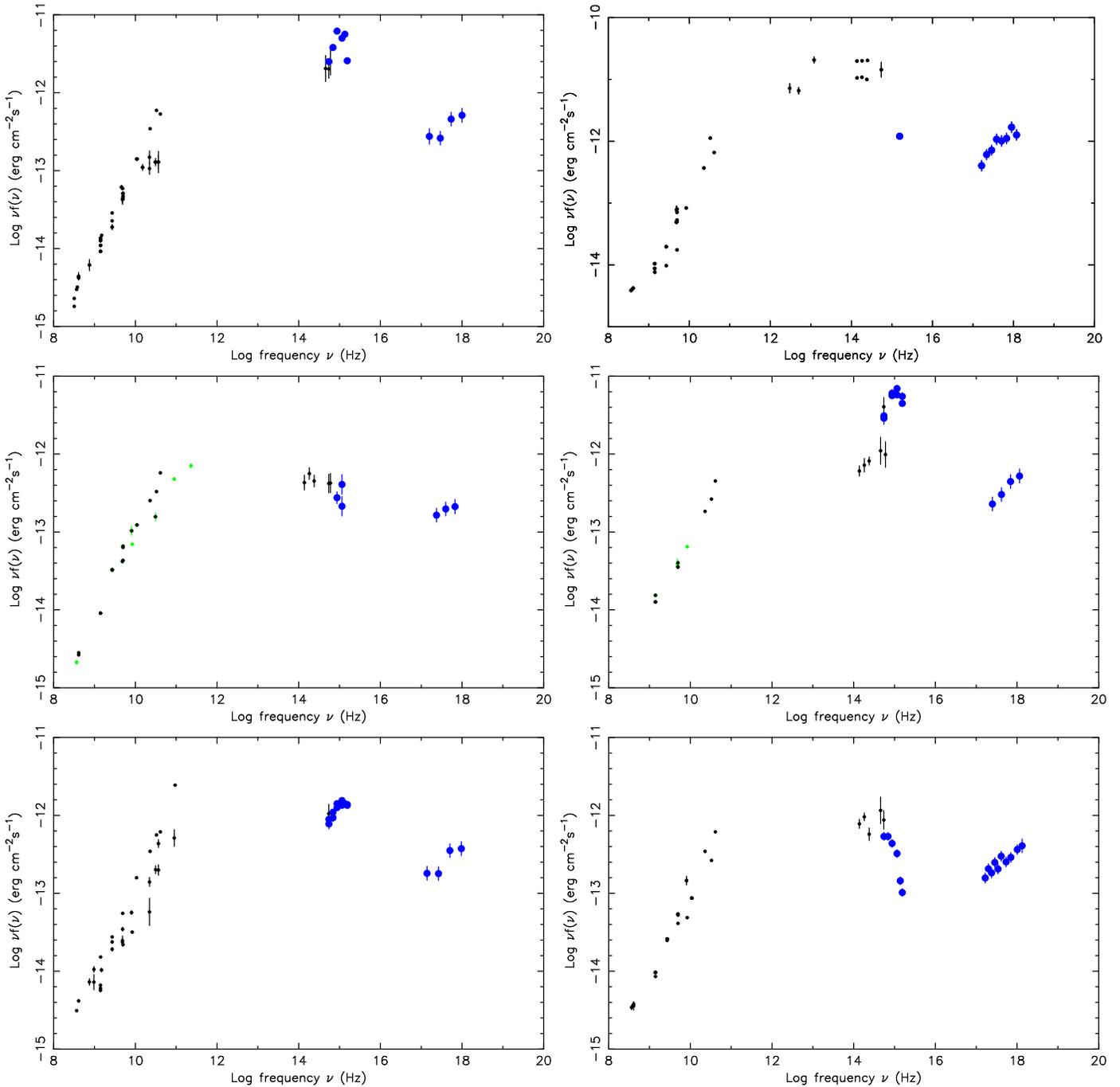

 \begin{center}
\vbox{
 \hbox{
 \epsfig{file=4160fig2.ps,width=6.cm,angle=-90}
 \epsfig{file=4160fig3.ps,width=6.cm,angle=-90}
 }
 \hbox{
  \epsfig{file=4160fig4.ps,width=6.cm,angle=-90}
  \epsfig{file=4160fig5.ps,width=6.cm,angle=-90}
 }
 \hbox{
 \epsfig{file=4160fig6.ps,width=6.cm,angle=-90}
 \epsfig{file=4160fig7.ps,width=6.cm,angle=-90}
 }
}
 \end{center}
\caption{Top left. The Spectral Energy Distribution of the blazar DW~0202+31=WMAP085.
Top right. The Spectral Energy Distribution of the blazar PKS 0215+015=WMAP096.
Mid left. The Spectral Energy Distribution of the blazar PKS~0511-220=WMAP127.
Mid right. The Spectral Energy Distribution of the blazar S5~0633+73=WMAP087.
Bottom left. The Spectral Energy Distribution of the blazar 1Jy~1030+415=WMAP103.
Bottom right. The Spectral Energy Distribution of the blazar 1Jy1406-07=WMAP203.
Swift-XRT and UVOT data are shown as large filled circles. }
 \label{seds_1}
\end{figure*}

\begin{figure*}[ht]
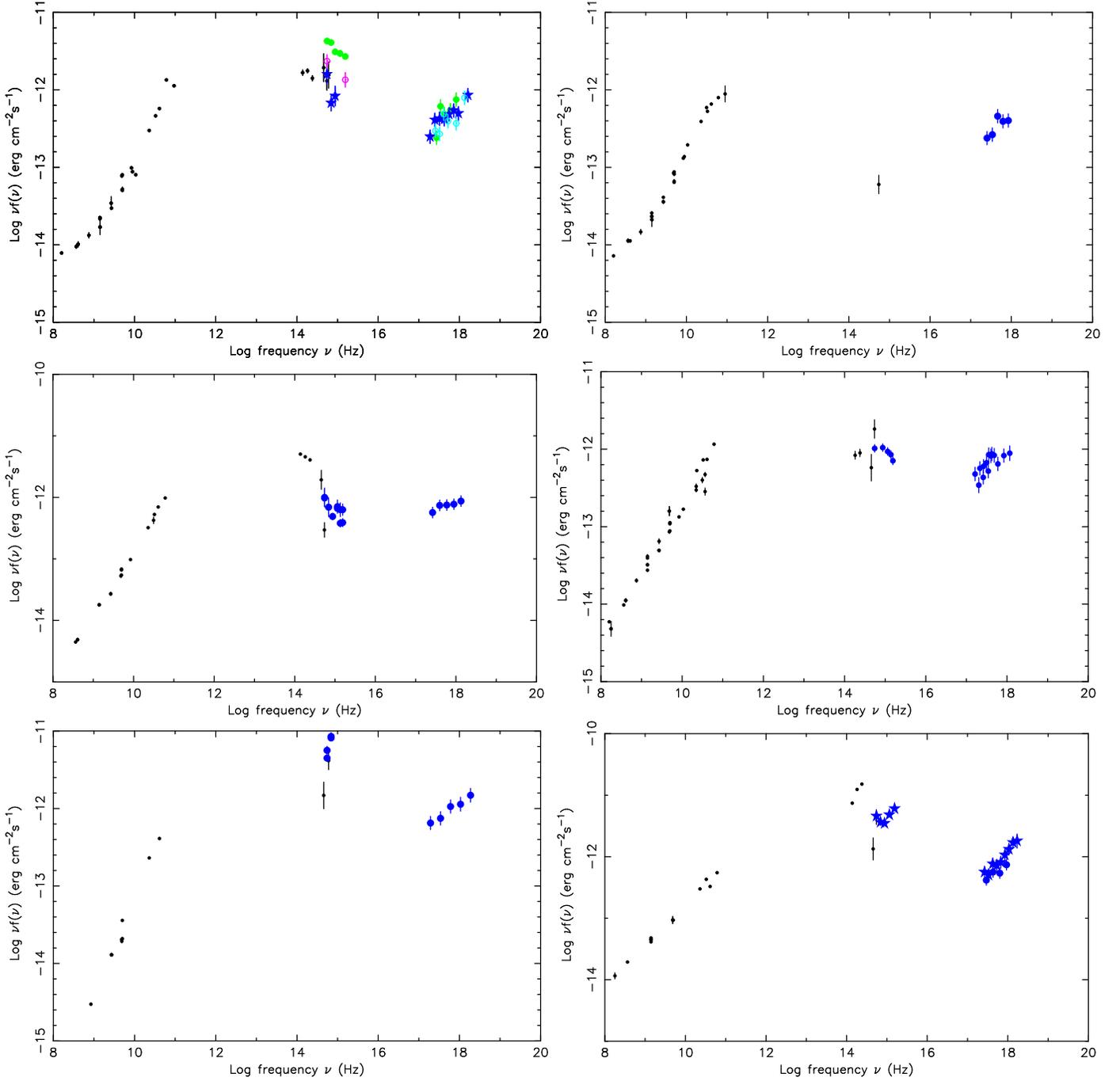

 \begin{center}
\vbox{
 \hbox{
 \epsfig{file=4160fig8.ps,width=6.cm,angle=-90}
 \epsfig{file=4160fig9.ps,width=6.cm,angle=-90}
 }
 \hbox{
 \epsfig{file=4160fig10.ps,width=6.cm,angle=-90}
 \epsfig{file=4160fig11.ps,width=6.cm,angle=-90}
 }
 \hbox{
 \epsfig{file=4160fig12.ps,width=6.cm,angle=-90}
 \epsfig{file=4160fig13.ps,width=6.cm,angle=-90}
 }
}
 \end{center}
\caption{Top left. The Spectral Energy Distribution of the blazars 1Jy0805-077=WMAP133.
 Swift-XRT are shown as filled stars (22 May 2005), open circles (4 June 2005) and filled circles (25 September 2005).
UVOT data were taken only on May 22 (filled stars), June 4 (open circles) and October 1
(filled circles). Note that while the optical/UV data (end of the synchrotron component)
shows significant variability, the X-ray data (inverse Compton component) are consistent
with a constant flux.
Top right. The Spectral Energy Distribution of the blazar 1Jy1213-172=WMAP173. Swift-XRT
data taken on 17 December 2005 are shown as large filled circles. No UVOT data are
available for this source because of the presence of a 2.8 magnitude star (HD106625) in
the UVOT field of view.
Mid left. The Spectral Energy Distribution of the blazar PKS 1313-333=WMAP182. Swift-XRT
and UVOT data taken on 7 September 2005 are shown as large filled circles.
Mid right. The Spectral Energy Distribution of the blazar 1Jy1548+056=WMAP007. Swift-XRT
and UVOT data are shown as large filled circles.
Bottom left. The Spectral Energy Distribution of the blazar PKS~1725-795=WMAP186.
Swift-XRT and UVOT data are shown as large filled circles.
Bottom right. The Spectral Energy Distribution of the blazar 3C 395=WMAP034. Swift-XRT
(Windowed timing) and UVOT data are shown as large star symbols and large filled circles
(Photon Counting data).
 } \label{seds_2}
\end{figure*}

\begin{figure*}[ht]
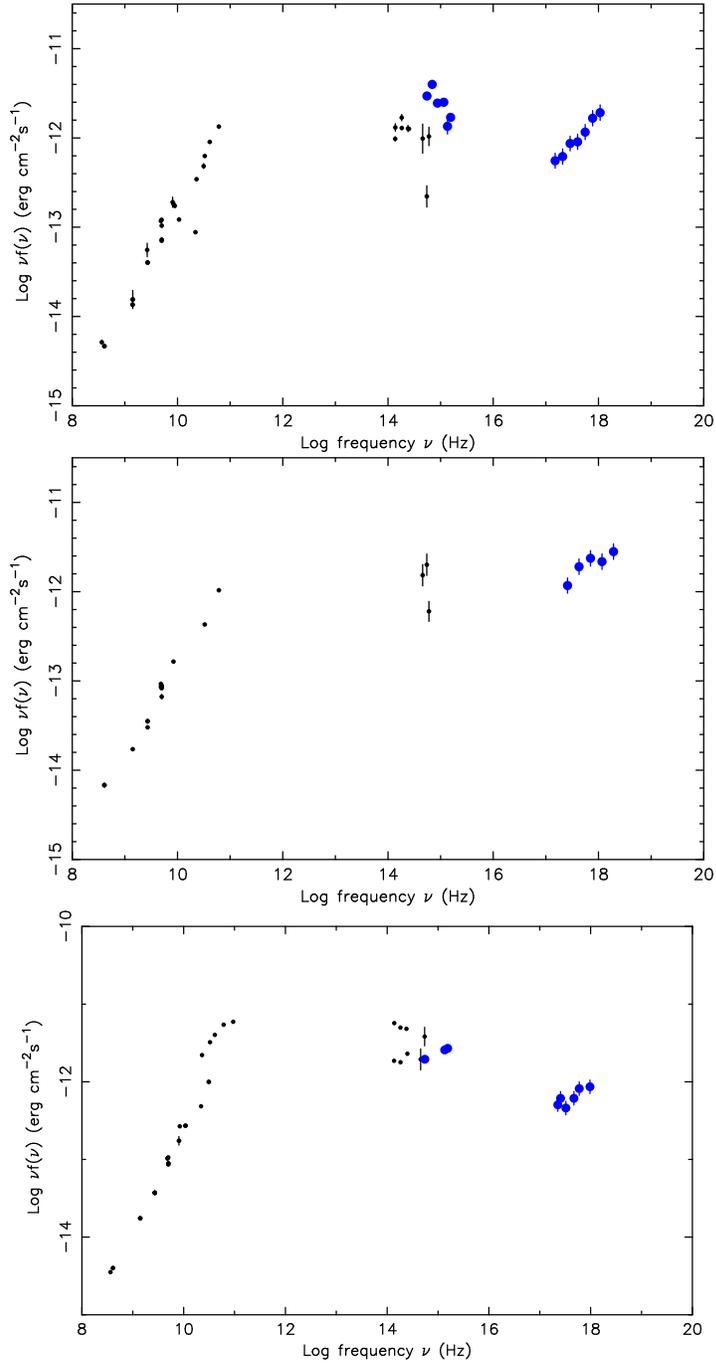

 \begin{center}
\vbox{
 \epsfig{file=4160fig14.ps,width=6.cm,angle=-90}
 \epsfig{file=4160fig15.ps,width=6.cm,angle=-90}
 \epsfig{file=4160fig16.ps,width=6.cm,angle=-90}
 }
 \end{center}
\caption{Top. The Spectral Energy Distribution of the blazar 1Jy2227-088=WMAP024.
Swift-XRT and UVOT data are shown as large filled circles.
Mid. The Spectral Energy Distribution of PKS~2355$-$534=WMAP189. Swift-XRT and UVOT data
are shown as large filled circles.
Bottom. The Spectral Energy Distribution of 1Jy2255$-$282=WMAP012. Swift-XRT and UVOT
 data are shown as large filled circles.
 }
 \label{seds_3}
\end{figure*}

The X-ray spectral slopes of the objects for which spectral analysis was possible are
flat, typical of inverse Compton emission as can be seen from Tabs. \ref{tab.spectral}
and \ref{tab.HR} and from the SEDs shown in Figs. \ref{seds_1}, through \ref{seds_3}.
In fact, all the SEDs shown in Figs. \ref{seds_1} through \ref{seds_3} clearly show that
all our objects are of the LBL type with the Swift XRT X-ray data sampling the rise of
the Inverse Compton emission, while the UVOT data represent either the tail of the
synchrotron component or, in some cases, the UV blue bump.

We have also looked for time variability in our data since flux variations of the inverse
Compton component of blazars is not well known, as most of the blazars that have been
monitored at these frequencies are of the HBL type. We found that no rapid variability
was detected in any source.
For the sake of completeness, we mention that fairly large X-ray flux variations have
been, however, observed in the inverse Compton emission from e.g. 3C273 and 3C279
(\cite{Turner90,ballo02,Gio02c,Courv03}) but only in exposures separated by months or
years. Large variations (of a factor of a few) on time scales of days, were also recently
detected in the LBL blazar 3C 454.3 by the Swift XRT in a series of observations carried
out during a giant optical and X-ray flare (\cite{Gio06b,pian06}).

Since the detection of all the 23 objects considered in this work brings the detection
rate in the X-ray band to 100\% of the 140 WMAP confirmed blazars with 41 GHz flux larger
than 0.8 Jy, we can safely conclude that {\it  all micro-wave selected} blazars {\it are
X-ray sources} with radio to X-ray flux ratio similar to that of blazars selected at GHz
frequencies, and in general of LBL blazars.

The micro-wave to X-ray  spectral slopes  (\amx, see Eq. \ref{eq.alphamux}) of our
sources are all within the very tight range seen in the sample of WMAP sources considered
by \cite{Gio06a}. We therefore confirm that the \amx distribution of all WMAP blazars of
the LBL type ($\approx $85 \% of the entire population) is very narrow and, consequently,
that  the X-ray flux is a very good estimator of the micro-wave flux.

This result may also apply to radio galaxies as the average \amx of the 6 objects of this
type included both in the WMAP catalog and in the Rosat all sky survey is 1.04, very
similar to that of blazars ($\langle $ \amx $\rangle = 1.07$).  We note, however, that
the soft X-ray emission in radio galaxies is not necessarily dominated by inverse Compton
radiation as it may include other components like, e.g., radiation from an accretion
disk, etc. The \amx distribution of radio galaxies could therefore have a lower average
and, probably, a wider dispersion.

As shown by \cite{giocol04} and by \cite{Gio06a}, blazars are the main foreground
contaminants of CMB anisotropy maps and can alter the reconstruction of the CMB
fluctuation spectrum in a significant way. The  measurement of the X-ray flux of blazars
and radio galaxies over a large area of the sky would therefore be extremely useful to
locate a large number of these sources and to determine their clustering properties.
\cite{Gio06a} showed that a deep all sky survey with flux limit of the order of
$10^{-15}$ \ergs in the soft X-ray band would detect well over 100,000 blazars, a
database that could be used to perform a careful cleaning of high sensitivity CMB data
from Planck and future CMB missions. This data set would  also be ideally suited for a
detailed estimate of the spatial correlation function of blazars and radio galaxies; this
information could be used to infer the residual impact of even fainter objects of this
type on the CMB temperature and polarization fluctuation spectrum.

The observations presented here provide valuable optical/UV and X-ray data for most  WMAP
blazars that lacked this type of broad-band multi-frequency information. The WMAP
micro-wave flux-limited sample of blazars complements the few existing and statistically
well defined samples selected at other frequencies like e.g. the 1Jy BL Lac sample
(\cite{Sti91}), the EMSS (\cite{Rec00}), the {\it Einstein} Slew survey
(\cite{Elvis92,Perlman96}) etc. Complete/flux limited samples with extensive
multi-frequency data are very important  for the investigation of  blazar statistical
properties from different viewpoints: a particularly interesting and relatively new one
is  the $\gamma$-ray band that is about to be deeply explored by the upcoming AGILE and
GLAST missions.

%%%%%%%%%%%%%%%%%%%
%
%
%%%%%%%%%%%%%%%%%%%
\begin{acknowledgements}
 We acknowledge financial support by the Italian Space Agency (ASI) through grant I/R/039/04 and
 through funding of the ASI Science Data Center. The UCL-MSSL authors acknowledge support of
 PPARC.
 We wish to thank all members of the Swift team who made this very flexible mission possible.
 This work is partly based and on data taken from the NASA/IPAC Extragalactic Database (NED)
 and from the ASI Science Data Center (ASDC).
\end{acknowledgements}
%%%%%%%%%%%%%%%%%%%%%%%%%%%%%

\end{document}